\documentclass[aps,prl,preprint,groupedaddress]{revtex4-1}

\newcommand{\bea}{\begin{equation}}
\newcommand{\eea}{\end{equation}}
\newcommand{\ber}{\begin{eqnarray}}
\newcommand{\eer}{\end{eqnarray}}

\begin{document}


\title{Conformation dependent damping and generalization of fluctuation-dissipation relation}


\author{A. Bhattacharyay}
\email[]{a.bhattacharyay@iiserpune.ac.in}
\affiliation{Indian Institute of Science Education and research, Pune, India}




\date{\today}

\begin{abstract}
Damping on an object generally depends on its conformation (shape size etc.). We consider the Langevin dynamics of
a model system with a conformation dependent damping and generalize the fluctuation dissipation relation to fit in
such a situation. We derive equilibrium distribution function for such a case which converges to the standard Boltzmann
form at the limit of uniform damping. The results can have implications, in general, for barrier overcoming processes where
standard Boltzmann statistics is slow.
\end{abstract}


\maketitle

\section{Introduction}
The Langevin dynamics for equilibrium processes in general includes a uniform damping constant $\Gamma$ over space for a particle 
that bears a relationship with the strength of the Gaussian white noise given by the expression $\sqrt{2\Gamma K_B T}$, where, 
$K_B$ is Boltzmann-constant and $T$ is the temperature of the bath with which the particle is in equilibrium. In such a
modelling, the average stochastic force experienced by the particle under consideration has to be a function (as mentioned above) of the 
damping constant of the particle to maintain an energy balance over mesoscopic time scales in equilibrium. The other important thing
is that the $\Gamma$ is generally structure dependent. For example, two spheres of different stokes radii in equilibrium with the same heat bath
will have two different $\Gamma$ and would see the same temperature $T$ of the bath. Keeping this in mind, an object which can have
conformation fluctuations in contact with a heat bath can in general have a damping constant dependent on its conformational states.
In the following, we would do an analysis for this kind of a system with a generalization of the fluctuation-dissipation (FD) relation in order to ensure 
a stationary probability distribution in the conformational space of the system. This is a necessity to enforce no current condition in the particularly
inhomogeneous conformational space of the system. The equilibrium distribution that we get is a generalization of Boltzmann distribution function which
becomes standard Boltzmann distribution at the limit of constant damping or on an average over the whole conformation-space.

\section{Model}
We consider a dimer model somewhat similar to the one we considered in Ref.\cite{ari}, to show a special case of having directed transport
under equilibrium conditions. In fact, the directed transport under equilibrium conditions would also have a general proof as a byproduct of our present exercise. The one-dimensional model in the over-damped regime
reads as
\ber
\Gamma_1(x_1-x_2)\dot{x_1}=-\frac{\partial U(x_1-x_2)}{\partial x_1} + \Gamma_1(x_1-x_2)\sqrt{\frac{2K_BT}{<\Gamma_1(x_1-x_2)>}}\eta_1(t)\\
\Gamma_2(x_1-x_2)\dot{x_2}=-\frac{\partial U(x_1-x_2)}{\partial x_2} + \Gamma_2(x_1-x_2)\sqrt{\frac{2K_BT}{<\Gamma_2(x_1-x_2)>}}\eta_2(t).
\eer 
In the above expressions, $Z = x_1-x_2$ is the coordinate that captures various conformations of the system, $U(Z)$ is the internal field that
keeps particles talking (e.g. keeps them bound and takes care of the excluded volume interaction) and there is no external global field present. The noises have been taken explicitly in accordance with the
generalized FD relation we talked about where $<\Gamma_i(Z)>$ is the spatial average of the damping constant which can be evaluated in a self consistent manner. Note that, for a uniform $\Gamma$ we
will recover the standard FD relation and the form of the standard FD relation is also there globally in terms of the average $\Gamma(Z)$. $\eta_i$s are the Gaussian white noise of
unit strength. Our subsequent analysis would justify the reason for such a generalization of the FD relation.
\par
Moving to the conformation and centre of mass ($X$) coordinates we can rewrite the above model as
\ber
\dot{Z}=-\frac{\Gamma_1(Z)+\Gamma_2(Z)}{\Gamma_1(Z)\Gamma_2(Z)}\frac{\partial U(Z)}{\partial Z} + \sqrt{2K_BT}\left( \frac{\eta_1(t)}{\sqrt{<\Gamma_1(Z)>}}-\frac{\eta_2(t)}{\sqrt{<\Gamma_2(Z)>}}\right)\\
\dot{X}=-\frac{\Gamma_1(Z)-\Gamma_2(Z)}{2\Gamma_1(Z)\Gamma_2(Z)}\frac{\partial U(Z)}{\partial Z} + \sqrt{\frac{K_BT}{2}}\left( \frac{\eta_1(t)}{\sqrt{<\Gamma_1(Z)>}}+\frac{\eta_2(t)}{\sqrt{<\Gamma_2(Z)>}}\right)
\eer
Note that, the second moment of the noise terms ($\xi_Z$ and $\xi_X$) in the Eq.3 and 4 respectively are
\ber
<\xi_Z(t_1)\xi_Z(t_2)>=\frac{2K_BT(<\Gamma_1(Z)>+<\Gamma_2(Z)>)}{<\Gamma_1(Z)><\Gamma_2(Z)>}\delta_{t_1-t_2}\\
<\xi_X(t_1)\xi_X(t_2)>=\frac{K_BT(<\Gamma_1(Z)>+<\Gamma_2(Z)>)}{2<\Gamma_1(Z)><\Gamma_2(Z)>}\delta_{t_1-t_2}
\eer 
where, the first moments vanish following the properties of $\eta_i$s. There is also a cross correlation between $\xi_Z$ and $\xi_X$ which we do not require to consider here because of the decoupling of the $Z$ dynamics from that of $X$. Let us consider the short hand notations 
$$
\zeta_Z=\frac{\Gamma_1(Z)+\Gamma_2(Z)}{\Gamma_1(Z)\Gamma_2(Z)}
$$
and
$$
\zeta_X=\frac{\Gamma_1(Z)-\Gamma_2(Z)}{2\Gamma_1(Z)\Gamma_2(Z)}
$$
and $F(Z)=-\frac{\partial U(Z)}{\partial Z}$.
\par
Now, the system to have a stationary probability distribution $P(Z)$ that makes the average current in the (inhomogeneous) conformation space $<\dot{Z}>$ vanish, we get the condition 
$<\dot{Z}>=<\zeta_ZF(Z)>=0$.
This means,
\bea
<\dot{Z}>=\int_a^b{dZ\zeta_ZF(Z)P(Z)}=0,
\eea
where the integration limits are at the two extreme zeros of the $P(Z)$ beyond which the conformations are not visited. Let us impose a sufficient condition for this integral to vanish
namely 
\bea
\frac{dP(Z)}{d(Z)}=C\zeta_ZF(Z)P(Z),
\eea
which would actually solve to give us 
\bea
P(Z)=N\exp{(C\int{dZ\zeta_ZF(Z)})},
\eea 
where $C$ is a constant that would make the exponent dimensionless and we would find it out from the corresponding
Fokker-Planck (FP) equation.
\section{Fokker-Planck equation}
The FP equation for the Langevin dynamics of the conformation coordinate i.e.
\bea
\dot{Z}=\zeta_ZF(Z) + \xi_Z(t)
\eea
can be found out following the standard procedure mentioned in \cite{swa}, using the equation
\bea
\frac{\partial P(Z)}{\partial t} = -\frac{\partial}{\partial x}<{\delta(x-Z(t))\dot{Z}}>,
\eea
where the $<>$ indicates a noise average for the noise distribution $P[\xi_Z(t)]=\exp{(-\int_{t_0}^{t_f}{dt\frac{\xi_Z(t)^2}{4\Gamma^\prime K_BT}})}$, where $\Gamma^\prime=\frac{<\Gamma_1(Z)>+<\Gamma_2(Z)>}{<\Gamma_1(Z)><\Gamma_2(Z)>}$ is a constant. The FP equation that would result from the above Langevin dynamics would be of the form
\bea
\frac{\partial P(x,t)}{\partial t} = -\frac{\partial}{\partial x}(\zeta_xF(x)P(x,t))+\Gamma^\prime K_BT \frac{\partial^2}{\partial x^2}P(x,t).
\eea
The normalized steady state solution for this FP is easily identifiable as
\bea
P(x)=N\exp{(\frac{\int{dx\zeta_xF(x)}}{\Gamma^\prime K_BT})}.
\eea
Thus, we identify the constant in Eq.9 which makes the exponent non-dimensional as $C=1/\Gamma^\prime K_BT$. Note that, for a constant damping we immediately recover the standard Boltzmann distribution function from our general expression for $P(x)$.
\section{discussion}
The first consequence of having this probability distribution is, in general, $\dot{X}=<\zeta_XF(Z)>\neq 0$ (compare the forms of $\zeta_X$ and $\zeta_Z$ in terms of $\Gamma_i$s). Thus, such a system can have a uniform centra of mass motion in the homogeneous bath. The homogeneity of bath-space does allow this unlike the conformation space which is inhomogeneous due to the presence of the internal field. Generally, people deny the presence of a CM motion under equilibrium conditions by mixing up ideas of the current in the presence of a global symmetry breaking field which appears for a non-equilibrium steady state with associated entropy production. But, in homogeneous space a CM velocity will not be associated with any entropy production because its a system in neutral equilibrium where some initial velocity gained while equilibrating can, in general, be maintained if symmetry breaking of the internal forces by damping (external force) can balance the overall damping on the CM motion. A particular case has already been discussed in \cite{ari}.
\par
Let us take a standard form of Langevin dynamics for a unit mass particle (for simplicity) with the conformation dependent damping and generalized FD relation as
\bea
\ddot{x}=-\Gamma(x)\dot{x} -\frac{\partial U(x)}{\partial x} +\Gamma(x)\sqrt{\frac{2K_BT}{<\Gamma(x)>}}\eta{t}
\eea
The corresponding velocity distribution being $P(v(x))=N\exp{\frac{-v(x)^2}{2K_BT\Gamma(x)/<\Gamma(x)>}}$ which is similar in form to the expression we have taken in terms of $\zeta$s and $\Gamma^\prime$ (which are basically inverse damping), one can do a local velocity average to get $<v(x)^2>$ with the above mentioned distribution to convert the strength of the noise to get the relation
\bea
\Gamma(x)^2\frac{2K_BT}{<\Gamma(x)>}=\Gamma(x)<v(x)^2>,
\eea
which is nothing but a statement of maintenance of local balance of energy. This readily gives a justification of the generalization of the Maxwell distribution of the system in accordance with the the generalized FD relation. 
This also lends support to the generalization of the FD relation on other hand showing that it keeps the local energy balance intact. 
\par
Let us try to understand the most striking advantage of such a situation of having generalized distribution for conformation dependent damping. Its just evident from the very sight of the probability distribution that an energy barrier overcoming can be aided by a large local damping which advantage is absent in the standard Boltzmann distribution. May be such a mechanism is in use in the problem of protein folding! May be its in use in a whole lot of other biological and glassy systems to help overcome the energy barriers and the landscape is not actually as rugged as it appears from the simple view of the potential itself! In experiments, actually measured deviation of the local probability distributions of the various conformations from the standard Boltzmann one can verify this present approach of generalization of the theory which probably so far not been done because of over reliance on the FD relation with a constant damping.
\par
The generalization of the FD relation in the present context, which was initiated by getting a zero average velocity in the inhomogeneous conformation space of the system in order to attain an equilibrium distribution, indicates a generalization of the standard Maxwell-Boltzmann distributions. This is not surprising, because, any system in equilibrium or in other words having equilibrium fluctuations would encounter an equilibrium damping. So, no system in equilibrium with a heat bath is actually a Hamiltonian system in the microscopic sense. It appears Hamiltonian in the presence of FD relation and at the scales on which the FD relation is defined. Moreover, the standard derivation of the FP with a uniform damping constant or just straightforward replacement of $\Gamma$ by $\Gamma(x)$ (which does not ensure zero current in the present model considered) gives us the clue as to why it does not show up in the equilibrium probability distribution, because, it cancels out in the numerator and the denominator. This process gives us the Boltzmann distribution which is a consequence of considering a Hamiltonian flow in the phase space (consequence of Liouville's equation). But, a nonuniform damping, which most generally can be the case, actually calls for a generalization and in the present work we have captured such a situation by suitably generalizing the FD relation. 
\par
Now, this generalization also imposes a problem. The damping now is not a constant only dependent upon the unknown degrees of freedoms that steal energy from the degrees of freedom are actually considered, rather, it being as well a function of the explicitly considered degrees of freedom there should be a prescription for finding it, otherwise, we lose predictability. This is where one should now look at to identify some general guiding principles for the conformation dependence of the damping where the unknown degrees of freedom part can feature as an adjustable constant.


\begin{thebibliography}{99}
\bibitem{ari}  A. Bhattacharyay, Physica A. {\bf 391}, 1111 (2012)
\bibitem{swa}  F. Schwabl, Statistical Mechanics, Springer International Edition, page-405 (chap.8)
\end{thebibliography}

\end{document}